\documentclass[aps,prl,twocolumn,groupedaddress,showpacs]{revtex4-1}

\usepackage{mathrsfs}
\usepackage{amsmath}
\usepackage{amssymb}
\usepackage{bm}
\usepackage{graphicx}
\usepackage{color}

\begin{document}


\title{Optical conductivity and surface plasmon modes of $U(1)$ spin liquid states with large spinon Fermi surfaces}


\author{Yuan-Fei Ma}
\author{Tai-Kai Ng}
\affiliation{Department of Physics, Hong Kong University of Science and Technology, Clear Water Bay Road, Kowloon, Hong Kong, China}


\date{\today}

\begin{abstract}
In this paper we study the optical properties of $U(1)$ spin liquids with large spinon Fermi surfaces based on a simple formula for the bulk optical conductivity obtained through the Ioffe-Larkin composition rule. We show that the optical conductivity of $U(1)$ spin liquids at energies above the charge gap has a unique feature that distinguishes them from ordinary insulators. In particular we show the existence of a long-life surface plasmon mode propagating along the interface between a linear medium and the spin liquid at frequencies above the charge gap, which can be detected by the widely used Kretschmann-Raether three-layer configuration.
\end{abstract}

\pacs{}

\maketitle

\allowdisplaybreaks
\newcommand{\ud}{\mathrm{d}}

\section{}

Recent experimental works on the organic compound $\kappa$-${\rm (ET)_2Cu_2(CN)_3}$\cite{Shimizu2003, Kezsmarki2006, Kurosaki2005, Yamashita2008a, Yamashita2008}, Herbertsmithite\cite{Helton2007, Han2012, Pilon2013} and other candidate materials\cite{Okamoto2007, Zhou2011, Itou2008} have pointed to the existence of the $U(1)$ quantum spin liquid state in nature. Spin liquids are Mott insulators with odd number of $1/2$-spin per unit cell which show no long-range magnetic order\cite{Lee2008b, Balents2010}. For the typical case of $\kappa$-${\rm (ET)_2Cu_2(CN)_3}$, the insulator is near the Mott metal-insulator transition and shows $U(1)$ spin liquid behavior including the existence of linear temperature dependence of the specific heat, the Pauli-like spin susceptibility and the nearly unity Wilson ratio, etc.

The low-energy effective theory of $U(1)$ spin liquids is believed to be described by a model composed of separate charge dynamics part $\mathcal{L}_c$ and spin dynamics part $\mathcal{L}_s$ (spin-charge separation). The charge dynamics $\mathcal{L}_c$ is described by the strong coupling phase of a quantum $x$-$y$ model in which the charge excitations acquire a gap, while the spin dynamics $\mathcal{L}_s$ has a free hopping Hamiltonian with a filled Fermi sea of spinons. The key feature of $U(1)$ spin liquids is that both parts are coupled to an emergent internal gauge field $a^{\mu} = (a_0, \mathbf{a})$, which leads to unusual properties of the spin liquids system\cite{Lee2005, Ng2007}. For example, although the spinons are neutral and do not couple to the external gauge field directly, the induced gauge field $a^{\mu}$ indirectly couples the external gauge field to the gapless spinons, yielding a power-law optical conductivity inside the Mott gap\cite{Ng2007}.

The optical conductivity plays an important role in the understanding of properties of ordinary metals and insulators, and we expect that it will also provide important information for the exotic spin liquid states. For the $U(1)$ spin liquids described by the low-energy effective $U(1)$ gauge theory $\mathcal{L}_s + \mathcal{L}_c$, the physical conductivity $\sigma (q, \omega)$ is related to the response functions of the spin and charge components through the Ioffe-Larkin composition rule\cite{Ioffe1989}:
\begin{eqnarray}
\label{eq1}
  \sigma^{-1}_{\alpha}(q, \omega) = \sigma^{-1}_{s \alpha}(q, \omega) + \sigma^{-1}_{c \alpha}(q,\omega),
\end{eqnarray}
where $s$ and $c$ denote the conductivities of the spin and charge components respectively, and $\alpha = l, t$ correspond to longitudinal and transverse components. In the $q\rightarrow0$ limit, there is no distinction between the longitudinal and transverse responses in an insulator, and we can replace the charge response simply by an effective dielectric function $\varepsilon_c(\omega)$. Then the physical optical conductivity can be written as\cite{Ng2007}:
\begin{eqnarray}
  \sigma(\omega) = \frac{\omega \sigma_{s}(\omega)}{ \omega + i \beta(\omega) \sigma_{s}(\omega)}, \label{eq:conductivity}
\end{eqnarray}
where $\beta(\omega) \equiv 4\pi/ (\varepsilon_c(\omega) - 1)$. The usual metallic state is recovered when $\beta(\omega) \rightarrow 0$ where the spinon conductivity becomes the physical conductivity. The spin liquid state is characterized by $\beta(\omega\rightarrow 0)>0$.
The unusual features of spin liquid states with large spinon fermi surface can be seen easily by using the Drude form for $\sigma_s(\omega) = \sigma_0/ (1 - i\omega \tau)$\cite{Ng2007}, where $\sigma_0=n e^2\tau / m$. Using Eq.\ (\ref{eq:conductivity}), we obtain
\begin{subequations}
\begin{eqnarray}
  \sigma(\omega) = \frac{\omega\sigma_0 }{ (1-i\omega \tau )\omega + i \beta(\omega) \sigma_0}, \label{eq:conductivity1}
\end{eqnarray}
in particular,
\begin{eqnarray}
   \mathrm{Re}\, \sigma(\omega)\rightarrow \frac{\omega^2}{\beta(0)^2\sigma_0} \sim \omega^2\tau^{-1},
\end{eqnarray}
\end{subequations}
in the limit $\omega \rightarrow 0$.

Generally, $\tau = \tau(\omega, T)$, where $\tau^{-1} \sim [\mathrm{max}(\hbar \omega, k_B T)]^{\alpha}$ at low temperature, leading to the power-law physical conductivity $\mathrm{Re} [\sigma(\omega)] \sim \omega^{2+\alpha}$ in the Mott gap, where $\alpha\sim1.33$ in gauge theory. The predicted in-gap power-law optical conductivity has indeed been observed in spin-liquid materials but with a different exponent $\alpha$ from the value proposed in gauge theory\cite{Pilon2013,Elsasser2012}.

In this paper we study the optical properties of $U(1)$ spin liquid states at energy range $\omega>\omega_g$, where $\omega_g$ is the charge gap which will be defined in the following. We shall show that the optical property of $U(1)$ spin liquid states with large fermi surfaces shows also strong deviation from ordinary insulators at this energy range. The difference can be seen most easily in the clean limit $\tau\rightarrow\infty$, where $\sigma(\omega)$ becomes
\begin{eqnarray}
  \sigma(\omega) \rightarrow \frac{i \omega \bar{\omega}_P^2} { \omega^2-\beta(\omega)\bar{\omega}_P^2}, \label{eq:conductivity2}
\end{eqnarray}
where $\bar{\omega}_P^2=\omega_P^2/4\pi$, $\omega_P^2=4\pi n e^2/m$ is the plasma frequency. Notice that
\[
   \mathrm{Re}\, \sigma(\omega)=\frac{\omega_P^2}{8} \left(\delta(\omega+\omega_g)+\delta(\omega-\omega_g)\right),
\]
and the $\delta$-function Drude weight of the clean metal at zero frequency is moved to a finite frequency $\omega_g=\sqrt{\beta(\omega_g)}\bar{\omega}_P$ in the spin liquid state, indicating that $\omega_g$ is the charge gap. The existence of a $\delta$-function Drude weight at $\omega_g$ is a unique feature of spin liquid states with large spinon fermi-surface and is consistent with the physical picture that spin liquid states are results of spin-charge separation at energies $\omega<\omega_g$\cite{Zhou2013a}. For $\omega>\omega_g$, spin and charge recombine. The $\delta$-function Drude weight at energy $\omega_g$ indicates that all spinons at the spinon fermi surface are all recombined into normal quasi-particles at this energy scale. This singular behavior is absent in ordinary insulators where conduction behavior is built up by exciting particles gradually and no singularity exists at the band edge.

At finite $\tau$, the $\delta$-function broadens into a (non-Lorentzian) peak centered at $\omega_g$. The total spectral weight under the peak is $\omega_P^2/8$, which is half of the spectral weight under the Drude peak in the corresponding metal state.
\begin{figure}
\includegraphics[width = 0.45 \textwidth]{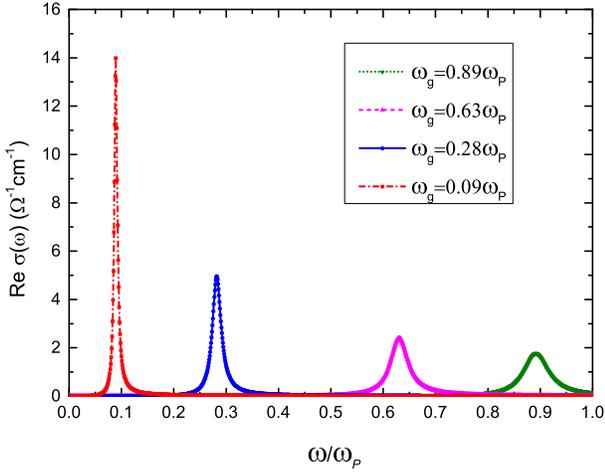}
\caption{\label{fig:sigma} Real part of the conductivity of the $U(1)$ spin liquids for different values of $\omega_g$. We choose $\tau^{-1}=0.1 \omega$ in our calculation.}
\end{figure}
This is shown schematically in Fig.~\ref{fig:sigma}, where we plot the real part of conductivity as a function of $\omega$ for several values of $\omega_g$. We have taken a $\omega$-independent $\beta(\omega)=\beta_0$ and $\tau^{-1}=0.1\omega$ in our calculation. The major effect of a $\omega$-dependent $\beta(\omega)$ is to renormalize $\omega_P$ and $\omega_g$ which will be explained later in the paper. Notice that the peaks gradually broaden as $\omega_g$ increases because of the increase in $\tau^{-1}$ as $\omega$ increases.

To see other consequences of this non-trivial behavior we consider the dielectric function given by
$\varepsilon(q, \omega) \equiv 1 + 4\pi i \sigma_l (q, \omega) /\omega$. Using Eq.\ (\ref{eq:conductivity1}) for conductivity, we obtain
\begin{subequations}
\label{dielectricspin}
\begin{eqnarray}
\mathrm{Re}\, \varepsilon (\omega) &=& 1 + \frac{4\pi \sigma_0 \tau (\omega^2_g -\omega^2)}{\omega^2 + \tau^2 (\omega^2_g -\omega^2 )^2},  \label{dielectricspinR}\\
\mathrm{Im}\, \varepsilon (\omega) &=&  \frac{4\pi \sigma_0 \omega }{\omega^2 + \tau^2 (\omega^2_g -\omega^2 )^2}. \label{dielectricspinI}
\end{eqnarray}
\end{subequations}

In ordinary metals, the dielectric function is in general complex, indicating that electromagnetic waves will decay inside the metal. However, it may still support surface modes propagating along the interface, say $x=0$, between a linear dielectric medium and the metal. We shall show that the same phenomenon exists in $U(1)$ spin liquid states at energies $\omega>\omega_g$. To be specific we consider the $p$-polarization case $\mathbf{H} = H \hat{z}$, the solutions (the TM modes) of the Maxwell equations that are wavelike in the $y$ direction can be written as
\begin{eqnarray}
  \mathbf{H}(\omega) &=& (0, 0, H) e^{i k y + i k_1 x }, \nonumber \\
  \mathbf{E}(\omega) &=& H \frac{c}{\omega} \frac{1}{\varepsilon_0} (-k, k_1, 0)e^{i k y + i k_1 x },\nonumber
\end{eqnarray}
in the region $x<0$ (the linear medium with dielectric constant $\varepsilon_0$), and as
\begin{eqnarray}
  \mathbf{H}(\omega) &=& (0, 0, H) e^{i k y + i k_{1s} x }, \nonumber \\
  \mathbf{E}(\omega) &=& H \frac{c}{\omega} \frac{1}{\varepsilon(\omega)} (-k, k_{1s}, 0)e^{i k y + i k_{1s} x },\nonumber
\end{eqnarray}
in the region $x>0$ (the spin liquids). Matching the boundary conditions gives the wave vectors\cite{raether1988surface, Pitarke2007}:
\begin{subequations}
\label{eq:wavevectors}
\begin{eqnarray}
  k^2 &=& \frac{\omega^2}{c^2} \frac{\varepsilon_0 \varepsilon(\omega)}{\varepsilon_0 +  \varepsilon(\omega)}, \\
  k_1^2 &=& \frac{\omega^2}{c^2} \frac{\varepsilon^2_0 }{\varepsilon_0 +  \varepsilon(\omega)}, \\
  k_{1s}^2 &=& \frac{\omega^2}{c^2} \frac{\varepsilon^2(\omega)}{\varepsilon_0 +  \varepsilon(\omega)}.
\end{eqnarray}
\end{subequations}

For the surface modes, the waves should be localized near the boundary and decay exponentially away from the boundary, while along the boundary, the waves should not decay. That is, the imaginary part of the wave vector $k_1$ should be negative, while the imaginary part of the wave vector $k_{1s}$ should be positive. So we can obtain the necessary conditions for the existence of the surface modes:
\begin{eqnarray}
\label{conditions}
 \mathrm{Im} k=0, \, \mathrm{Re} k_{1} \ll -\mathrm{Im} k_{1},\, \mathrm{Re} k_{1s} \ll \mathrm{Im} k_{1s}.
\end{eqnarray}

In the limit when the imaginary part of the dielectric function is small and can be neglected, these conditions can be satisfied in frequency ranges where the real part of the dielectric function $\varepsilon(\omega)$ is negative and $|\varepsilon(\omega)|>\varepsilon_0$. For ordinary metals this happens at frequency range $0<\omega<\omega_P/\sqrt{1+\varepsilon_0}$. To see what happens in the $U(1)$ spin liquid states we examine the $\tau\rightarrow\infty$ limit of the dielectric function given by Eqs.~(\ref{dielectricspin}). It is easy to see that $\mathrm{Im}\, \varepsilon(\omega)\rightarrow 0$ in this limit and
\[
\mathrm{Re}\, \varepsilon(\omega)\rightarrow 1- \frac{\omega_P^2} {\omega^2-\omega_g^2},\]
and conditions\ (\ref{conditions}) are satisfied for $\omega_g<\omega<\sqrt{\omega_P^2/(1+\varepsilon_0)+\omega_g^2}$ and we expect that surface plasmon mode exists in
this frequency range.

\begin{figure}
\includegraphics[width = 0.45 \textwidth]{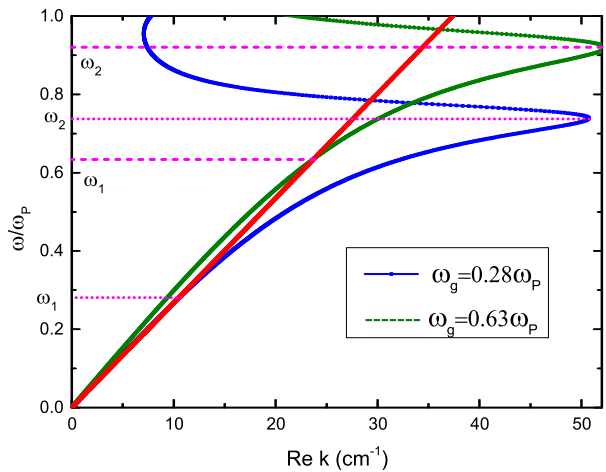}
\caption{\label{fig:komega} Energy spectral of the free photon (red line) $\omega = c k$, and the surface plasmon mode $\omega = \omega(k)$ for different values of $\omega_g$. In the region $\omega_1 < \omega < \omega_2$ the real part of $k$ is much larger than the imaginary part, and the real part of the dielectric function $\varepsilon$ is negative. It is easy to see as $\omega_g$ becomes larger, the frequency range becomes narrower.}
\end{figure}

The main effect of nonzero (but small) $\tau^{-1}$ is to introduce a small $\mathrm{Im} k\neq 0$, i.e. the surface plasmons acquire a finite propagation length and life time.
In this case, we may replace the plasmon existence condition $\mathrm{Im} k=0$ by a less demanding condition
\[
\mathrm{Re} k \gg \mathrm{Im} k.\]
The surface plasmon dispersion is computed with this modified criteria for several $\beta$ and are shown in Fig.~\ref{fig:komega} where we take $\tau^{-1}=0.1\omega$. We find that surface plasmon exists between frequencies $\omega_1\sim\omega_g$ and $\omega_2\sim\sqrt{\omega_P^2/(1+\varepsilon_0)+\omega_g^2}$ and the range between $\omega_1$ and $\omega_2$ narrows as $\beta$ increases because of the increase in $\tau^{-1}$ as a function of frequency.

For the organic material $\kappa$-${\rm (ET)_2Cu_2(CN)_3}$ with exchange coupling $J \approx 250 \mathrm{K} \sim 10 \mathrm{m} \mathrm{eV} \sim \epsilon_F$, using the relation $n/m = N(0) v^2_F/ d$ ($d$ is the dimension of the system), it is easy to estimate the frequency scale $\bar{\omega}_P = \sqrt{n e^2 / m} = e \sqrt{2 N(0) \epsilon_F /(d m)} \sim 10^{12} \mathrm{s}^{-1}$, or in inverse length unit $\bar{\omega}_P / (2 \pi c)  \sim 5.3 \,\mathrm{cm}^{-1}$, which lies in the low frequency regime\cite{Elsasser2012}. Notice the long length scale ($\sim$ cm) justifies the neglectance of $q$ dependence in $\varepsilon(\omega)$.

\begin{figure}
\includegraphics[width = 0.45 \textwidth]{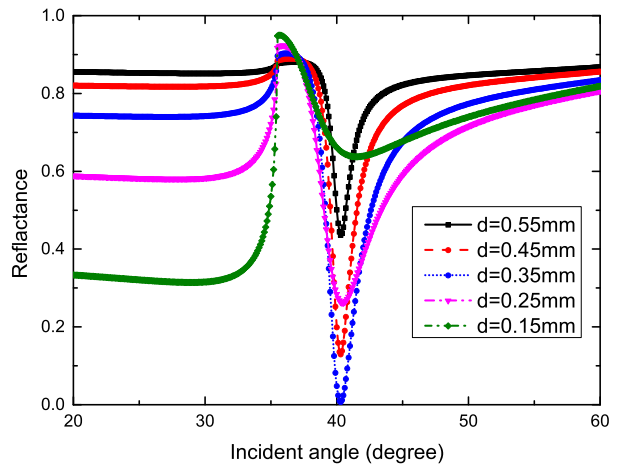}
\caption{\label{fig:R1}Reflectance at spin liquids film of different thickness in the Kretschmann-Raether configuration, the dip of the total reflectance corresponds to the resonant angle where the surface modes are exited. We can see that at a certain thickness $d=0.35$ mm, the incident waves are totally absorbed to excite the surface modes. The resonant angle is $\theta \approx 40^{\circ}$.}
\end{figure}

The surface modes cannot be excited by a single free-propagating photon because in the region $\omega_1 < \omega < \omega_2$ the dispersion curve of the surface modes lies always on the right of the free photon line, then various alternative excitation schemes have been developed. The most widely used scheme relies on the coupling via evanescent waves through the Kretschmann-Raether configuration\cite{Kretschmann1968, raether1988surface} using a prism-spin liquids-linear medium $(1|2|3)$ three-layer model. Then the total reflectance of the three-layer model is given by:
\[
  R = \left| \frac{r_{12} + r_{23} e^{2i k_{1s} d}}{1 + r_{12} r_{23} e^{2i k_{1s} d}} \right|^2, \]
where $d$ is the thickness of the medium 2 film (the spin liquids), and $r_{ij}$ is the reflection coefficient at the interface between layer $i$ and layer $j$:
\[
  r_{ij} = \frac{n_j \cos{\theta_i} - n_i \cos{\theta_j}}{n_j \cos{\theta_i} + n_i \cos{\theta_j}} = \frac{k_{i,x}/\varepsilon_i - k_{j,x}/\varepsilon_j }{k_{i,x}/\varepsilon_i + k_{j,x}/\varepsilon_j}, \]
where $\theta_i$ is the incident angle on the $i$th layer. The surface modes excitation is associated with characteristic dip in the total reflectance as shown in Fig.~\ref{fig:R1}. The refractive index of the prim is chosen as $n_1 = 1.73$, the units and values of the parameters used are summarized in Table~\ref{tab:parameter}.

\begin{table}
\caption{\label{tab:parameter}The units and values of the parameters used in the calculation.}
\begin{ruledtabular}
\begin{tabular}{ccccccc}
\textrm{parameter}& $\omega$ & $\bar{\omega}_P$ & $a$\footnotemark[1] & $\beta$ & $k$ & $d$ \\
\hline
\textrm{unit} & $10^{12}\mathrm{s}^{-1}$ & $10^{12}\mathrm{s}^{-1}$ & $-$  & $-$ & $\mathrm{cm}^{-1}$ & $\mathrm{mm}$ \\
\textrm{value} & 0.55\footnotemark[2] & 0.316 & .1  & 1 & $-$ & $0.35$\\
\end{tabular}
\end{ruledtabular}
\footnotetext[1]{Here we choose $\tau^{-1} = a \omega$, which is different from the one given in gauge theory.}
\footnotetext[2]{Frequency of the incident light to excite the surface modes, $\omega = 0.5 \omega_P$.}
\end{table}

\begin{figure}
\includegraphics[width = 0.45 \textwidth]{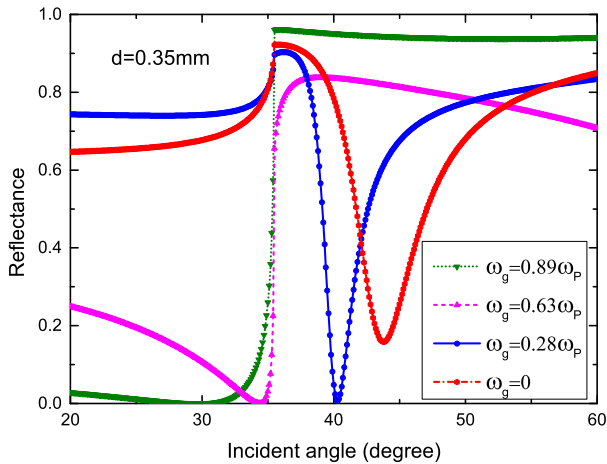}
\caption{\label{fig:R2}Reflectance at spin liquids film for different values of $\omega_g$ at the thickness $d=0.35$ mm. The dip in reflectance gradually smoothes out and disappears when $\omega_g$ becomes large, indicating that the system behaves more and more like an ordinary insulator.}
\end{figure}

The reflectance for fixed thickness $d=0.35$mm as a function of $\omega_g$ is shown in Fig.~\ref{fig:R2}, where we see that the dip in reflectance gradually smoothes out and disappears when $\omega_g$ becomes large, indicating that the system behaves more and more like an ordinary insulator.

We now consider the $\omega$-dependence of $\beta(\omega)$. We note from Eq.\ (\ref{eq1}) that
\[
   -\frac{i\omega}{\beta(\omega)}=\sigma_c(\omega)\sim \frac{i\omega \omega_c^2}{\omega^2-m^2}.  \]
The last equality gives the general form of $\sigma_c(\omega)$ computed from the strong coupling phase of $\mathcal{L}_c$. $\omega_c^2$ measures the spectral weight of the slave rotors and $m$ is the rotors' charge gap. Therefore, $\beta(\omega)\sim \omega_c^{-2}(m^2-\omega^2)$. Putting this back into Eq.\ (\ref{eq:conductivity2}) we see that the frequency-dependence of $\beta(\omega)$ renormalized $\omega^2_P \rightarrow\omega_R^{2}$ with
\[
\omega^{-2}_R=\omega_P^{-2}+\omega_c^{-2},\]
and $\omega_g=m(\omega_R/\omega_c)$, which are results of the Ioffe-Larkin composition rule. In particular, the qualitative conclusions we drawn above remain unchanged.

Lastly we apply our analysis to Herbertsmithite which is believed to be a $U(1)$ spin liquid with Dirac fermion like dispersion\cite{Pilon2013,Potter2013}. In this case it is expected that
\begin{equation}
\label{diracc}
\sigma_s(\omega)\sim \frac{e^2}{8}=\sigma_D,
\end{equation}
which is an universal number independent of disorder\cite{Potter2013}. Notice that no Drude-like peak exists in this case. Putting in Eq.\ (\ref{eq:conductivity}), we obtain $\sigma(\omega)\sim \omega^2$ at low frequency\cite{Potter2013}. The corresponding dielectric function is
\begin{subequations}
\label{dielectricspinD}
\begin{eqnarray}
\mathrm{Re}\, \varepsilon (\omega) &=& 1 + \frac{4\pi \sigma_D^2\omega_c^{-2}(m^2-\omega^2)}{\omega^2 + \sigma_D^2\omega_c^{-4}(m^2-\omega^2)^2},  \label{dielectricspinDR}\\
\mathrm{Im}\, \varepsilon (\omega) &=&  \frac{4\pi \omega\sigma_D }{\omega^2 + \sigma_D^2\omega_c^{-4}(m^2-\omega^2)^2}, \label{dielectricspinDI}
\end{eqnarray}
\end{subequations}
which has the same form as Eqs.\ (\ref{dielectricspin}) if we replace $\tau\rightarrow\sigma_D\omega_c^{-2}$, $\sigma_0\rightarrow\sigma_D$ and $\omega_g\rightarrow m$. With realistic parameters, we estimate that the effective $\tau$ is too small and $\mathrm{Re}\, \varepsilon (\omega)$ stays positive for all frequencies $\omega$ suggesting that no surface plasmon mode exists in this case.

Summarizing, we discuss in this paper optical properties of $U(1)$ spin liquids with large spinon Fermi surfaces based on the conductivity obtained from gauge theory and Ioffe-Larkin rule. We show how the charge gap $\omega_g$ arises in the optical conductivity through the parameter $\beta(\omega)$ and that the $\delta$-function Drude weight in the metallic state is transferred to finite frequency $\omega_g$ in the Mott insulating state.

This in terms leads to the prediction that surface plasmon modes propagating along the interface between a linear medium and a $U(1)$ spin liquid with large spinon fermi surface are supported at frequency $\omega>\omega_g$, the surface mode can be excited by the widely used Kretschmann-Raether three-layer configuration. On the other hand, surface plasmon is absent in Herbertsmithite which is believed to be a $U(1)$ spin liquid with Dirac fermion like dispersion.

\acknowledgments
  We acknowledge support from HKRGC through grant No. 603913.
\subsection{}
\subsubsection{}

\appendix
\section{}

\begin{acknowledgments}
\end{acknowledgments}

\bibliography{SL}

\end{document}